# Phosphorylation by the stress-activated MAPK Slt2 down-regulates the yeast TOR complex 2

Kristin L. Leskoske,[1,3,6] Françoise M. Roelants,[1,6] Anita Emmerstorfer-Augustin,[1,4] Christoph M. Augustin,[2,5] Edward P. Si,[1] Jennifer M. Hill,[1] and Jeremy Thorner[1]

[1]Department of Molecular and Cell Biology, [2]Department of Mechanical Engineering, University of California at Berkeley, Berkeley, California 94720, USA

*Saccharomyces cerevisiae* target of rapamycin (TOR) complex 2 (TORC2) is an essential regulator of plasma membrane lipid and protein homeostasis. How TORC2 activity is modulated in response to changes in the status of the cell envelope is unclear. Here we document that TORC2 subunit Avo2 is a direct target of Slt2, the mitogen-activated protein kinase (MAPK) of the cell wall integrity pathway. Activation of Slt2 by overexpression of a constitutively active allele of an upstream Slt2 activator (Pkc1) or by auxin-induced degradation of a negative Slt2 regulator (Sln1) caused hyperphosphorylation of Avo2 at its MAPK phosphoacceptor sites in a Slt2-dependent manner and diminished TORC2-mediated phosphorylation of its major downstream effector, protein kinase Ypk1. Deletion of Avo2 or expression of a phosphomimetic Avo2 allele rendered cells sensitive to two stresses (myriocin treatment and elevated exogenous acetic acid) that the cell requires Ypk1 activation by TORC2 to survive. Thus, Avo2 is necessary for optimal TORC2 activity, and Slt2-mediated phosphorylation of Avo2 down-regulates TORC2 signaling. Compared with wild-type Avo2, phosphomimetic Avo2 shows significant displacement from the plasma membrane, suggesting that Slt2 inhibits TORC2 by promoting Avo2 dissociation. Our findings are the first demonstration that TORC2 function is regulated by MAPK-mediated phosphorylation.

Cellular homeostasis requires careful coordination of metabolic growth processes and adaptive stress responses. Multiple signaling networks coordinate expansion at the cell boundary with the accumulation of intracellular mass and progression through the cell cycle and also play critical roles in responding to environmental fluctuations that compromise the integrity of the plasma membrane (PM) and, in yeast, the cell wall (Levin 2011).

Like all eukaryotes, yeast contains two evolutionarily conserved multicomponent protein kinase complexes in which a TOR (target of rapamycin) polypeptide is the catalytic subunit: TOR complex 1 (TORC1) and TORC2. TORC1 and TORC2 are essential regulators of growth and homeostasis but control different aspects of yeast cell physiology (González and Hall 2017; Tatebe and Shiozaki 2017). TORC2 itself localizes to the PM (Berchtold and Walther 2009). TORC2 is thought to monitor the condition of the cell envelope (PM and cell wall) and control, through downstream effectors, processes that maintain the integrity and function of these vital barriers. The primary target of TORC2 is the AGC family protein kinase Ypk1 (and its paralog, Ypk2) (Kamada et al. 2005; Roelants et al. 2011; Berchtold et al. 2012; Niles et al. 2012). TORC2 activates Ypk1 by phosphorylating its C-terminal segment at multiple sites, two of which are highly conserved (the so-called "turn" and "hydrophobic" motifs), but all of which are required for maximal Ypk1 catalytic activity and for Ypk1 stability (Leskoske et al. 2017). Ypk1 in turn phosphorylates multiple substrates that modulate PM lipid and protein composition (for review, see Gaubitz et al. 2016; Roelants et al. 2017a). It also has been reported (Nomura and Inoue 2015) that TORC2 phosphorylates C-terminal turn-like and hydrophobic-like motifs in Pkc1, an AGC protein kinase that controls

Present addresses: [3]Collaborative Center for Translational Mass Spectrometry, Translational Genomics Research Institute, Phoenix, AZ 85004, USA; [4]Institute of Molecular Biotechnology, Graz University of Technology, Graz 8010, Austria; [5]Institute of Biophysics, Medical University of Graz, Graz 8010, Austria.
[6]These authors contributed equally to this work.
Corresponding author: jthorner@berkeley.edu

the cell wall integrity (CWI) pathway (Levin 2011); however, whether TORC2-mediated phosphorylation of Pkc1 affects its localization, stability, or activity has not been demonstrated.

Conditions that perturb the PM stimulate TORC2-mediated phosphorylation of Ypk1, including sphingolipid depletion (Roelants et al. 2011), hypotonic stress (Berchtold et al. 2012; Niles et al. 2012), heat shock (Sun et al. 2012), and elevated exogenous acetic acid (Guerreiro et al. 2016), whereas hyperosmotic stress inhibits TORC2-mediated phosphorylation of Ypk1 (Lee et al. 2012; Muir et al. 2015). How TORC2 "senses" different stresses could, in some cases, involve direct feedback control. For example, sphingolipids seem to negatively regulate TORC2 because treatment of cells with a potent inhibitor (myriocin/ISP-1) of L-serine:palmitoyl-CoA C-palmitoyltransferase, the enzyme that catalyzes the first committed step in sphingolipid biosynthesis, causes up-regulation of TORC2-dependent and Ypk1-mediated phosphorylation of proteins whose modification markedly stimulates the rate of sphingolipid production (Roelants et al. 2011; Muir et al. 2014).

In addition, however, other pathways that sense PM or cell wall stress may exert some of their effects via regulation of TORC2 function. In this regard, two mitogen-activated protein kinase (MAPK) pathways in yeast sense and respond to stresses that challenge the cell envelope: the high-osmolarity glycerol (HOG) pathway, whose terminal MAPK is Hog1 (Saito and Posas 2012), and the CWI pathway, whose terminal MAPK is Slt2/Mpk1 (Levin 2011). Hog1 activation, which can occur via a branch dependent on PM sensor Sln1 or a branch dependent on PM sensor Sho1, is required for cell survival under hypertonic conditions (Saito and Posas 2012) but can also be evoked by other environmental insults, including both heat shock (Winkler et al. 2002) and cold shock (Panadero et al. 2006), citric acid stress (Lawrence et al. 2004), hypoxia (Hickman et al. 2011), glucose starvation (Vallejo and Mayinger 2015), and, most notably from the perspective of this study, sphingolipid depletion (Tanigawa et al. 2012). Slt2 activation, which is initiated via stretch sensors anchored in both the PM and the cell wall (Wsc1, Mid2, and others) (Levin 2011), is required for cell survival under hypotonic conditions (Davenport et al. 1995) but can also be elicited by heat shock (Kamada et al. 1995) and agents that directly damage the cell wall, such as zymolyase, Calcofluor White, or Congo Red (de Nobel et al. 2000; García et al. 2004; Rodríguez-Peña et al. 2013). When activated, the stretch sensors recruit to the PM and activate Rom2 and/or Tus1—guanine nucleotide exchange factors (GEFs) for the Rho1 GTPase. GTP-bound Rho1 then binds to and activates Pkc1, which in turn activates the CWI MAPK cascade (Levin 2011). Rho1 and its GEFs also localize at sites of polarized growth to promote cell wall glucan synthesis, actin cytoskeleton organization, and targeting of secretory vesicles to growth sites (Perez and Rincón 2010).

Significant cross-talk exists between the HOG and CWI pathways (Hahn and Thiele 2002; García-Rodríguez et al. 2005; Fuchs and Mylonakis 2009; García et al. 2009) and between these MAPK pathways and TORC2 function. For example, down-regulation of TORC2–Ypk1 signaling and activation of Hog1 act in concert to maintain cell viability during hyperosmotic stress (Muir et al. 2015). In addition, one of the first functions reported for Tor2 (the TOR subunit unique to TORC2) was proper polarization of the actin cytoskeleton (Schmidt et al. 1996), linking Tor2 to Rho1 and the CWI pathway. Indeed, various means to stimulate CWI pathway output rescue the inviability and actin polarization defects associated with loss of TORC2 or Ypk1 function (Helliwell et al. 1998; Roelants et al. 2002). In addition, TORC2–Ypk1 signaling, through its role in regulating lipid composition and organization, influences PM localization of Rho1 and Rom2 (Niles and Powers 2014; Hatakeyama et al. 2017).

In this study, we sought to establish whether the CWI MAPK pathway directly modulates TORC2–Ypk1 signaling, given the cumulative evidence for the interconnections between them. As documented here, we found that activation of the CWI pathway MAPK results in direct Slt2-mediated phosphorylation of TORC2 subunits and that these modifications down-regulate TORC2–Ypk1 signaling. By pinpointing the molecular basis of this negative regulation, our results provide new mechanistic insight into how TORC2 senses cell envelope stress.

## Results

### TORC2 subunits are targets of the Slt2 MAPK

TORC2 is a 1.4-MDa dimeric complex in which each protomer is a hetero-oligomer comprising Tor2, Lst8, Avo1, Avo2, Avo3, and either Bit2 or Bit61 (Wullschleger et al. 2005; Karuppasamy et al. 2017). Phosphoproteomic analyses (for compilations, see the *Saccharomyces* Genome Database, https://www.yeastgenome.org) have detected in vivo phosphorylation of -SP- and -TP- sites in four of the six TORC2 subunits (Fig. 1A). Most strikingly, nine such sites are present in Avo2, and four confirmed sites cluster at its C-terminal end; 11 sites are present in Avo3, and five confirmed sites cluster at its N-terminal end (Fig. 1A). Both Hog1 and Slt2 phosphorylate only -SP- or -TP- motifs (Mok et al. 2010), just like the proline-directed Ser- or Thr-specific MAPKs in other eukaryotes (Roux and Blenis 2004).

To determine whether any -SP- or -TP- sites in TORC2 subunits are direct targets of the Slt2 MAPK, we focused initially on Avo2. As expected for a phosphoprotein, Avo2 exists, even under basal conditions, in a series of isoforms of distinct mobility, as resolved by Phos-tag SDS-PAGE (Fig. 1B), a technique in which the degree of retardation of a protein reflects the extent of its phosphorylation (Kinoshita et al. 2009). A convenient method to activate Slt2 in the absence of an external stimulus is overexpression of a constitutively active Pkc1 allele, Pkc1[R398A R405A K406A] (referred to here as *PKC1\**) (Martín et al. 2000), which is hyperactive because residues critical for the inhibitory function of its pseudosubstrate sequence have been mutated. After induction of *PKC1\** expression, Avo2 became hyperphosphorylated, and the increase in Avo2 phosphorylation coincided with the kinetics of appearance of active Slt2 (Fig. 1B). Overexpression of a catalytically inactive ("kinase-dead" [KD]) derivative, Pkc1*(D949A), to the same level as Pkc1* (Fig. 1C) did not activate Slt2 or alter Avo2 phosphorylation (Fig. 1B).

In addition to robust Slt2 activation, we noted that overexpression of Pkc1* also evoked Hog1 activation, in agreement with other observations demonstrating cross-talk between these two MAPK pathways (Hahn and Thiele 2002; García-Rodríguez et al. 2005; Fuchs and Mylonakis 2009; García et al. 2009). Therefore, to distinguish which MAPK was primarily responsible for the Avo2 hyperphosphorylation induced by Pkc1* overexpression, we repeated these experiments in both *slt2Δ* and *hog1Δ* mutants. We found that the Pkc1*-induced mobility shift was markedly reduced in *slt2Δ* cells but not in *hog1Δ* cells (Fig. 1D). Moreover, to confirm that the Pkc1*-induced

Slt2-mediated modifications were occurring on the -SP- and -TP- sites, the same experiments were repeated with an Avo2 allele (Avo2^9A) in which all nine sites were mutated to Ala. Indeed, absence of these sites prevented Pkc1*-induced hyperphosphorylation of Avo2 (Fig. 1D). Although Slt2 appears to be the major MAPK responsible for Pkc1*-induced hyperphosphorylation of Avo2, these modifications were not completely abrogated in *slt2Δ* cells and were modestly reduced in *hog1Δ* cells compared with wild-type cells, suggesting that Hog1 makes a minor contribution to Pkc1*-induced hyperphosphorylation of Avo2.

In further analysis of the capacity of Slt2 to modify Avo2, wild-type Slt2 immuno-enriched from cells overexpressing Pkc1*—but not a catalytically deficient derivative, Slt2(K54R), purified in the same manner—robustly phosphorylated recombinant wild-type Avo2 in vitro,

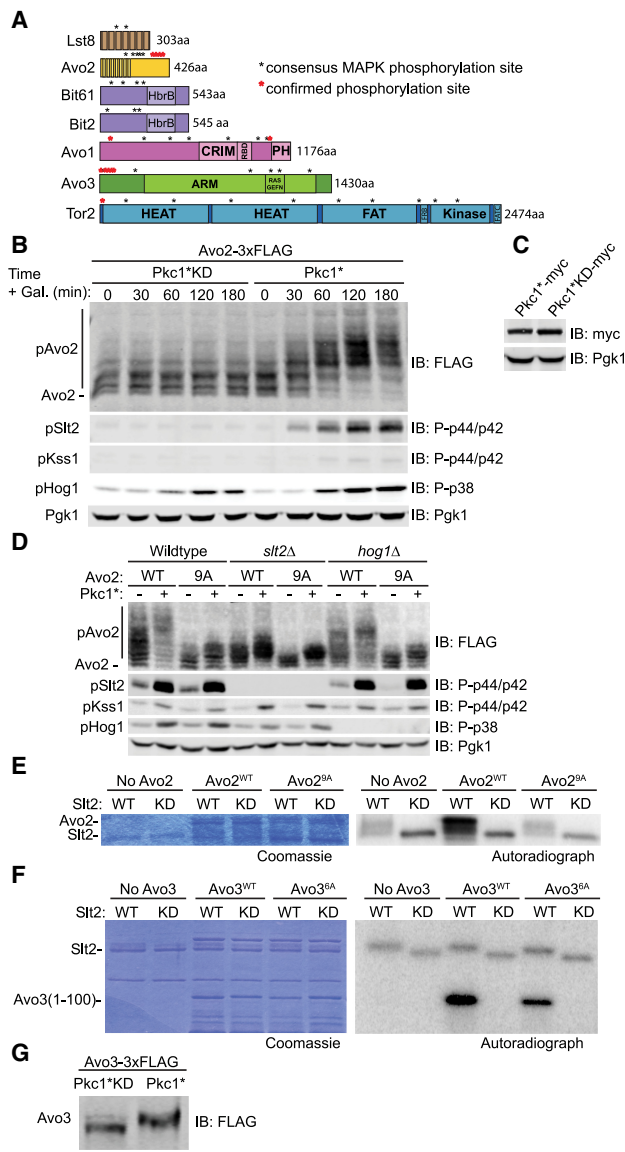

**Figure 1.** MAPK Slt2 phosphorylates TORC2 subunits Avo2 and Avo3. (*A*) The primary structure of each indicated TORC2 subunit is depicted schematically, with domains labeled as in Gaubitz et al. (2016). (Black asterisk) An -SP- or -TP- site; (red asterisk) an -SP- and -TP- detectably phosphorylated in vivo in various phosphoproteomic analyses, as cataloged in the *Saccharomyces* Genome Database (http://www.yeastgenome.org). (*B*) Wild-type cells (JTY5336) carrying both a *CEN* plasmid expressing Avo2-3xFlag (pKL1) from the *AVO2* promoter and a multicopy (2 μm DNA) vector expressing either Pkc1* (pJT5660) or catalytically inactive Pkc1*KD (pJEN12) from the *GAL1* promoter were cultured to mid-exponential phase in selective minimal medium containing 2% raffinose and 0.2% sucrose. Expression of Pkc1* or Pkc1*KD was induced by addition of galactose (2% final concentration). Cell samples were removed at the indicated times and lysed, the proteins in the resulting extracts were resolved by Phos-tag SDS-PAGE, and the indicated proteins were analyzed by immunoblotting with the appropriate antibodies, all as described in the Materials and Methods. Pgk1 was the loading control. (*C*) As in *B* except the cells carried plasmids expressing either Pkc1*-myc (pAEA376) or Pkc1*KD-myc (pJEN13), and the extracts were resolved by standard SDS-PAGE prior to immunoblotting. (*D*) Wild-type cells (JTY5336) or otherwise isogenic *slt2Δ* (YFR549) or *hog1Δ* (YFR538-A) derivatives carrying plasmids expressing either Avo2-3xFlag (pKL1) or Avo2^9A-3xFlag (pKL2), as indicated, as well as either empty vector (yEplac112) or the same vector expressing Pkc1* from the *GAL1* promoter (pJT5660), were grown to mid-exponential phase in selective minimal medium containing 2% raffinose and 0.2% sucrose. After addition of galactose (2% final concentration), the cells were cultured for 3 h, harvested, and lysed, and the indicated proteins were analyzed as in *B*. (*E*) Activated wild-type Slt2 (pKL63) and kinase-dead (KD) Slt2 (pKL64) were purified from *Saccharomyces cerevisiae* as described in the Materials and Methods and incubated with [γ-32P]ATP and either GST-Avo2^WT (pKL16) or GST-Avo2^9A (pKL17) purified from *Escherichia coli* as described in the Materials and Methods. Reaction products were resolved by SDS-PAGE and analyzed by Coomassie blue staining (*left*) and autoradiography (*right*). (*F*) As in *E*, except the substrates were purified recombinant GST-Avo3(1-100) (pKL81) or GST-Avo3(1-100)^6A (pKL82). (*G*) A strain (YFR617) expressing Avo3-3xFlag and carrying a plasmid expressing either Pkc1*KD (pJEN12) or Pkc1* (pJT5660) from the *GAL1* promoter was grown, induced with galactose for 2.5 h, lysed, and analyzed by Phos-tag SDS-PAGE as in *B*.

and this incorporation was eliminated in an Avo2⁹ᴬ mutant (Fig. 1E).

Likewise, Avo3(1–100), an N-terminal fragment containing six of the 11 -SP- and -TP- sites in Avo3 (five of which have been detected as phosphorylated in vivo), was robustly phosphorylated in vitro by wild-type Slt2 but not by the catalytically deficient Slt2 mutant, and incorporation was markedly reduced by mutation of the six sites to Ala (Fig. 1F). Moreover, as judged by its mobility shift, intact Avo3-3xFlag was phosphorylated in vivo upon overexpression of Pkc1* but not upon overexpression of Pkc1*KD (Fig. 1G). Collectively, these results show that at least two core subunits of TORC2 are physiologically relevant substrates of the Slt2 MAPK.

## MAPK phosphorylation of Avo2 impairs TORC2 activity

Unlike Avo3 (1430 residues), which is essential for TORC2 function because it has an important structural role (Wullschleger et al. 2005; Karuppasamy et al. 2017), Avo2 is a relatively small (426 residues) peripherally located nonessential subunit of TORC2, with the highest overall density of -SP- and -TP- sites (Fig. 1A), suggesting that its Slt2-mediated phosphorylation might have a readily detectable regulatory role. To examine this possibility, in addition to wild-type Avo2 and the MAPK site-deficient Avo2⁹ᴬ mutant, we also generated a phosphomimetic allele, Avo2⁹ᴱ. First, we found that all three proteins were expressed at comparable steady-state levels (Fig. 2A); thus, the absence or presence of MAPK phosphorylation does not affect Avo2 stability. Strikingly, however, even though Avo2 is nonessential for cell viability under nonstress conditions, we found that Avo2 is essential for cell survival during myriocin-induced sphingolipid depletion (Fig. 2B, top row), a condition in which cells require full TORC2-mediated activation of Ypk1 to survive (Roelants et al. 2011; Leskoske et al. 2017). Furthermore, compared with cells expressing wild-type Avo2 or Avo2⁹ᴬ, cells expressing Avo2⁹ᴱ were much more sensitive to the growth inhibitory effect of myriocin (Fig. 2B), suggesting that, like avo2Δ cells, they are deficient in TORC2-dependent Ypk1 activation. In direct support of that conclusion, the myriocin sensitivity of both the avo2Δ mutant and Avo2⁹ᴱ-expressing cells was rescued by coexpression of a Ypk1 allele, Ypk1(D242A), that is fully active in the absence of TORC2 phosphorylation (Fig. 2C; Kamada et al. 2005; Roelants et al. 2011; Leskoske et al. 2017). These findings indicate that Avo2 is required for optimal TORC2 function and that phosphorylation interferes with Avo2 function, resulting in down-regulation of TORC2 activity.

These conclusions were confirmed by monitoring TORC2 activity in vivo. We documented previously that TORC2 phosphorylates Ypk1 at multiple sites that lie within 40 residues of its C terminus and that full phosphorylation of these sites is necessary for both maximal Ypk1 activity and Ypk1 stability (Leskoske et al. 2017). Therefore, to assess TORC2 activity in vivo, we examined phosphorylation of Ypk1 at four of its C-terminal TORC2-dependent phosphorylation sites using Phos-tag SDS-

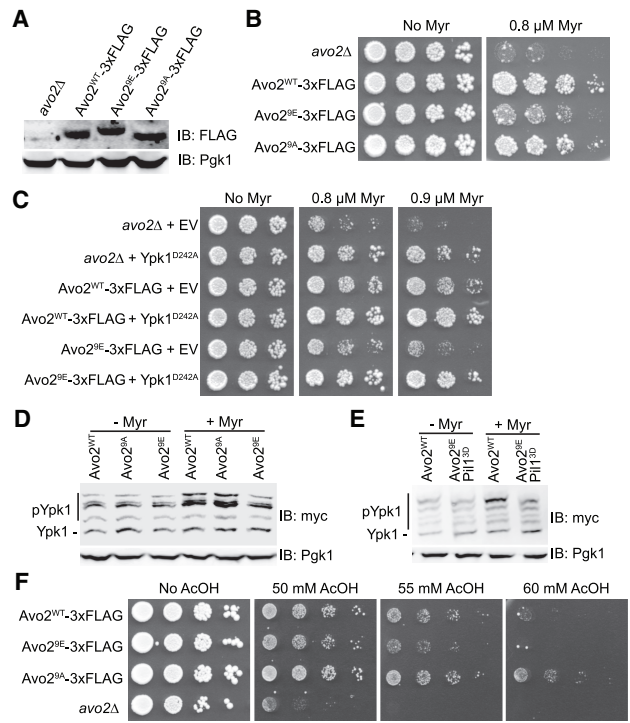

**Figure 2.** Phosphorylation of Avo2 at its MAPK sites impairs TORC2 activity. (A) Otherwise isogenic strains JTY7318 (avo2Δ), yKL32 (Avo2ᵂᵀ-3xFlag), yKL34 (Avo2⁹ᴱ-3xFlag), and yKL33 (Avo2⁹ᴬ-3xFlag) were grown to mid-exponential phase in rich medium, harvested, and lysed, and the resulting extracts were resolved by SDS-PAGE and analyzed by immunoblotting, as described in the Materials and Methods. Pgk1 was the loading control. (B) Overnight cultures of the strains in A were adjusted to $A_{600\ nm} = 0.1$, spotted undiluted and in a series of fivefold serial dilutions on YPD plates either lacking or containing 0.8 μM myriocin, incubated for 3 d at 30°C, and then imaged. (C) Strains JTY7318 (avo2Δ), yKL32 (Avo2ᵂᵀ-3xFlag), or yKL34 (Avo2⁹ᴱ-3xFlag) expressing either empty vector (pRS315) or Ypk1ᴰ²⁴²ᴬ-myc from the same vector (pFR234) were adjusted to $A_{600\ nm} = 0.1$, spotted as in B but on SCD-L-T plates lacking or containing myriocin at the indicated concentrations, and imaged for 3 d at 30°C. (D) Wild-type (BY4742), Avo2⁹ᴬ (YFR528), or Avo2⁹ᴱ (yKL31) strains expressing Ypk1⁵ᴬ-myc from its native promoter on a CEN plasmid (pFR246) were cultured to mid-exponential phase in selective minimal medium and then treated with vehicle (methanol) or 1.25 μM myriocin for 2 h. After harvesting, whole-cell lysates were prepared, resolved by SDS-PAGE, and analyzed as described in the Materials and Methods. A representative of three independent experiments is shown. (E) As in D, except strains BY4742 and Avo2⁹ᴱ Pil1³ᴰ (YFR641) were used. (F) Overnight cultures of the strains in A were adjusted to $A_{600\ nm} = 0.1$, spotted undiluted and in a series of 10-fold serial dilutions on SCD plates either lacking or containing acetic acid at the indicated concentrations, and imaged after incubation for 3 d at 30°C.

PAGE. Treatment with myriocin, a condition that activates TORC2 activity in wild-type cells, failed to stimulate TORC2 phosphorylation of Ypk1 in Avo2⁹ᴱ-expressing cells (Fig. 2D). As presented further below, Avo2 has also been found in association with PM structures called eisosomes (Douglas and Konopka 2014), and Pil1, a major eisosome component, has been shown to be another target of Slt2 (Mascaraque et al. 2013). Hence, we also examined Ypk1 phosphorylation in cells expressing both Avo2[9E] and Pil1[3D] and again found that, unlike in wild type or cells expressing Pil1[3D] alone (Supplemental Fig. S1), myriocin treatment failed to stimulate TORC2 phosphorylation of Ypk1 in the cells expressing both Avo2[9E] and Pil1[3D] (Fig. 2E). These additional observations verify that it is phosphorylation of Avo2 that makes a major contribution to interfering with TORC2 function.

Another stress that cells require TORC2-mediated Ypk1 activation to survive is exposure to elevated exogenous acetic acid (Guerreiro et al. 2016). As with myriocin treatment, avo2Δ cells were markedly more sensitive to this stress than wild-type cells, and cells expressing Avo2[9E] were detectably more sensitive than cells expressing either wild-type Avo2 or especially Avo2[9A] (Fig. 2F). These results again show that Avo2 is required for full TORC2 activity and that MAPK phosphorylation of Avo2 impairs TORC2 function.

## Slt2 MAPK action down-regulates TORC2 function

As assessed by Phos-tag SDS-PAGE, after induction of Pkc1*KD, there was no change in the pattern of TORC2-dependent Ypk1 isoforms over the course of 2.5 h, whereas after induction of Pkc1*, there was a decrease in the slowest mobility (most highly phosphorylated) isoforms, concomitant with the appearance of activated Slt2 (Fig. 3A), as well as a marked reduction of total Ypk1 protein (Fig. 3A,B), indicating that activation of the CWI pathway prevented TORC2-mediated phosphorylation of Ypk1. The observed Pkc1*-induced down-regulation of TORC2-mediated Ypk1 phosphorylation and loss of Ypk1 protein was largely prevented in a slt2Δ mutant but not in hog1Δ or kss1Δ single mutants or a hog1Δ kss1Δ double mutant (Fig. 3C). We examined a kss1Δ mutation alone or in combination with other MAPK-null mutations because we noted that overexpression of Pkc1* sometimes mildly increased the amount of dually phosphorylated (activated) Kss1 MAPK (Fig. 1D). However, compared with the slt2Δ cells, the levels of Ypk1 the loading control. (Red dots) TORC2-dependent isoforms of Ypk1 that persist. (D) As in C, except the cells coexpressed Pkc1*KD from the GAL1 promoter on the plasmid (pJEN12). (E) Wild-type (BY4741) cells expressing Ypk1[5A]-myc from its native promoter on a CEN plasmid (pFR246) were cultured to mid-exponential phase in selective minimal medium and then treated with increasing amounts of Congo Red (0, 10, 20, or 40 µg/mL final concentration) for 2 h. After harvesting, whole-cell lysates were prepared, resolved by SDS-PAGE, and analyzed as described in the Materials and Methods. (F) Same as in E, except that wild-type (JTY5336) and slt2Δ (YFR549) strains were used, and, when added, Congo Red was added at 20 µg/mL final concentration.

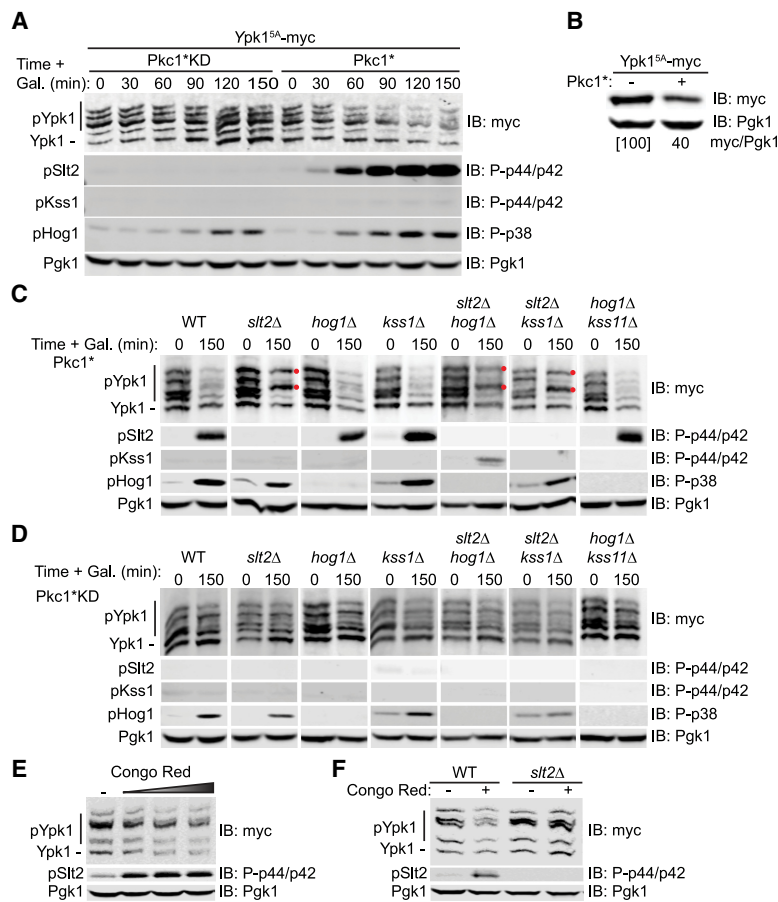

Figure 3. Activation of MAPK Slt2 down-regulates TORC2-mediated Ypk1 phosphorylation. (A) Strain JTY5336 expressing Ypk1[5A]-myc from its native promoter on a CEN plasmid (pFR246) and expressing from the GAL1 promoter on plasmids either Pkc1*KD (pJEN12) or Pkc1* (pJT5660), as indicated, were cultured to mid-exponential phase in selective minimal medium containing 2% raffinose and 0.2% sucrose and induced by addition of galactose (2% final concentration), and samples taken at the indicated time points were lysed and analyzed by Phos-tag SDS-PAGE and immunoblotting, as described in the Materials and Methods. Pgk1 was the loading control. (B) Strain JTY5336 expressing Ypk1[5A]-myc from its native promoter on a CEN plasmid (pFR246) and carrying either empty vector (−) (yEPlac112) or expressing Pkc1* (+) from the GAL1 promoter on the same vector were grown as in A and harvested 2 h after galactose induction, and the relative level of Ypk1 protein was determined and normalized to the Pgk1 loading control (ratio was set at 100 for the empty vector control) by SDS-PAGE and immunoblotting. A representative image is shown for an experiment that was repeated three independent times. (C) Strains JTY5336 (HOG1*KSS1*SLT2*), YFR549 (slt2Δ), YFR538-A (hog1Δ), YFR560 (kss1Δ), YFR559 (slt2Δ hog1Δ), YFR567 (slt2Δ kss1Δ), and YFR564 (hog1Δ kss1Δ) expressing Ypk1[5A]-myc from its native promoter on a CEN plasmid (pFR246) and coexpressing Pkc1* from the GAL1 promoter on a plasmid (pJT5660) were cultured to mid-exponential phase in selective minimal medium containing 2% raffinose and 0.2% sucrose and induced by addition of galactose (2% final concentration), and, after 2.5 h, samples were lysed and analyzed as in A. Pgk1 was phosphorylation or Ypk1 protein were not further enhanced in *hog1Δ slt2Δ* or *kss1Δ slt2Δ* double mutants (Fig. 3C). Most revealingly, in every case where Slt2 is absent, two of the most prominent and highly TORC2 phosphorylated Ypk1 isoforms persisted (Fig. 3C, red dots). Thus, Slt2 is the MAPK primarily responsible for inhibiting TORC2-mediated Ypk1 phosphorylation.

It has been reported that Pkc1 is a substrate of TORC2 (Nomura and Inoue 2015); hence, it was possible that overexpression of even Pkc1*KD might titrate away a sufficient amount of TORC2 to interfere with its efficient phosphorylation of Ypk1. However, this was not the case because, in either wild-type cells or all of these same mutants, induction of Pkc1*KD had little or no effect on the pattern of the Ypk1 isoforms (Fig. 3D). Furthermore, treatment of cells expressing normal Pkc1 and Slt2 at their endogenous levels with the cell wall-perturbing dye Congo Red, a physiological condition that has been demonstrated to elicit Pkc1-dependent Slt2 activation (de Nobel et al. 2000), resulted in down-regulation of TORC2 function (Fig. 3E), and the down-regulation was dependent on Slt2 (Fig. 3F).

## TORC2 is down-regulated during sustained Sln1 inactivation

Sln1 is a histidine kinase similar to those in bacterial two-component signaling systems, and its inactivation, as occurs during hyperosmotic shock, leads to activation of the Hog1 MAPK (Saito and Posas 2012). In a global screen of the *Saccharomyces cerevisiae* kinome, we found that, like a *tor2ts* mutant (Leskoske et al. 2017), when a *sln1ts* mutant was shifted to the restrictive temperature, there was a dramatic loss of TORC2-dependent Ypk1 phosphorylation (Supplemental Fig. S2A) that was accompanied by hyperphosphorylation of Avo2 (Supplemental Fig. S2B), indicating that, like activation of the CWI MAPK pathway, activation of the HOG MAPK pathway down-regulates TORC2 function and does so via a similar mechanism.

To confirm this conclusion and inactivate Sln1 independently of any external temperature or hypertonic stress, we constructed a strain expressing the plant F-box protein Tir1 and an allele of Sln1 tagged with a modified auxin-inducible degron (AID*), which should cause Sln1 to be degraded in an SCF-mediated manner upon auxin addition (Morawska and Ulrich 2013). Adding a synthetic auxin (1-naphthaleneacetic acid [1-NAA]) to cells expressing untagged Sln1 had no effect on the level of Sln1 or on the Ypk1 isoforms present and did not activate either Hog1 or Slt2 (Fig. 4A, left). The same treatment of cells expressing AID*-tagged Sln1 resulted in its rapid degradation, as expected, as well as the loss of the Ypk1 isoforms, concomitant with the activation of both Hog1 and Slt2 (Fig. 4A, right), in agreement with prior work showing that Hog1 activation leads to Slt2 activation (García-Rodríguez et al. 2005). Thus, we again found that MAPK activation coincided with down-regulation of TORC2 activity, as judged by the dramatic drop in Ypk1 C-terminal phosphorylation and protein level.

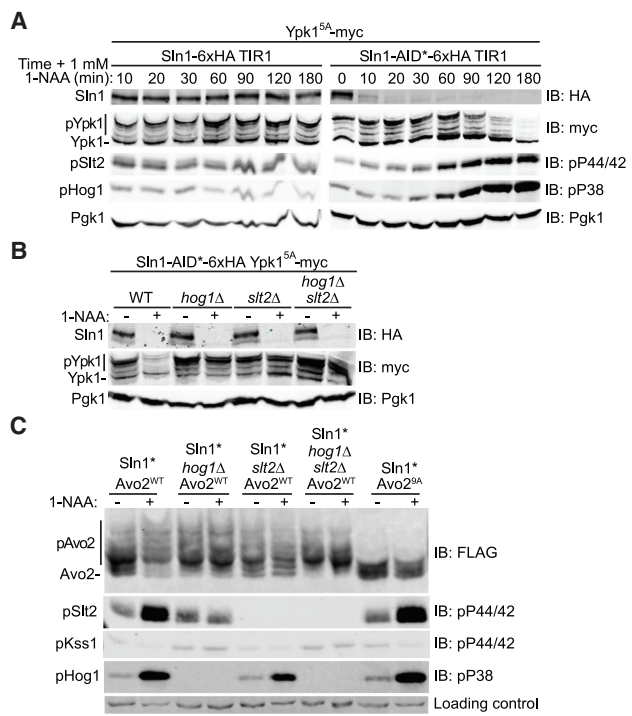

**Figure 4.** TORC2 function is down-regulated after Sln1 degradation. (*A*) Strains yKL15 (Sln1-6xHA TIR1) or yKL18 (Sln1-AID*-6xHA TIR1) expressing Ypk1[5A]-myc from its native promoter on a *CEN* plasmid (pFR246) were grown to mid-exponential phase in phosphate-buffered selective minimal medium (pH 6.2) and treated with 1 mM 1-NAA (final concentration), and samples were withdrawn at the indicated times, lysed, subjected to Phos-tag SDS-PAGE to resolve Ypk1 phosphorylation and to standard SDS-PAGE to resolve the other proteins, and analyzed by immunoblotting as described in the Materials and Methods. (*B*) Strains yKL18 (Sln1-AID*-6xHA TIR1 *HOG1* *SLT2*), yKL20 (Sln1-AID* TIR1 *hog1Δ*), yKL16 (Sln1-AID* TIR1 *slt2Δ*), and yKL22 (Sln1-AID* TIR1 *hog1Δ slt2Δ*), all expressing Ypk1[5A]-myc from its native promoter on a *CEN* plasmid (pFR246), were grown as in *A*, treated with either solvent alone (DMSO) or 1 mM 1-NAA (final concentration) in the same volume of solvent for 90 min, harvested, lysed, and analyzed as in *A*. (*C*) The same strains as in *B*, except expressing either Avo2[WT]-3xFlag (pKL1) or Avo2[9A]-3xFlag (pKL2), as indicated, were grown and treated as in *B*, except that the incubation with 1-NAA was for 2 h, and samples of each culture were analyzed as in *A*, with Avo2 phosphorylation resolved by Phos-tag SDS-PAGE, and the other proteins resolved by standard SDS-PAGE.

Consistent with the cross-talk by which activation of Hog1 leads to activation of Slt2 (García-Rodríguez et al. 2005), the drastic reduction in Ypk1 C-terminal phosphorylation and protein level caused by Sln1 degradation was ameliorated in either a *hog1Δ* mutant or a *slt2Δ* mutant and was not further enhanced in a *hog1Δ slt2Δ* double mutant (Fig. 4B). Moreover, similar to what we observed upon Pkc1*-induced CWI pathway activation, activation of the HOG pathway by Sln1 degradation markedly stimulated phosphorylation of Avo2, as judged by the shift to a spectrum of slower mobility isoforms in the Phos-tag gel (Fig. 4C, left two lanes), and no such shift was observed for the Avo2[9A] mutant, indicating that these modifications were occurring on its -SP- and -TP- sites (Fig. 4C, right two lanes). Furthermore, there was no change in the pattern of Avo2 isoforms upon Sln1 degradation in cells lacking either Hog1 or Slt2 (or both) (Fig. 4C, middle lanes), indicating that Avo2 phosphorylation was MAPK-dependent. Importantly, activation of Hog1 alone (upon degradation of Sln1 in $slt2\Delta$ cells) was not sufficient to down-regulate TORC2-mediated Ypk1 phosphorylation (Fig. 4B) or stimulate Avo2 phosphorylation (Fig. 4C), indicating that Slt2 is the MAPK responsible for these effects and that, in this case, the primary role of activated Hog1 is to stimulate production of activated Slt2.

Interestingly, we documented previously that when the HOG pathway is activated by exposure of cells to 1 M sorbitol, a cell-impermeable osmolyte that robustly but transiently activates Hog1 (Saito and Posas 2012), there is rapid and transient collapse of the TORC2-dependent Ypk1 isoforms, but that response still occurs even in $hog1\Delta$ cells (Lee et al. 2012; Muir et al. 2015). Although treatment with 1 M sorbitol activates Hog1, it does not activate Slt2, and the collapse of the TORC2-dependent Ypk1 isoforms still occurs in cells lacking either Slt2 or Hog1 (or both) (Supplemental Fig. S3). Thus, in contrast to the Slt2-dependent down-regulation of TORC2-mediated Ypk1 phosphorylation that occurs upon prolonged inactivation of Sln1, the loss of the TORC2-dependent Ypk1 isoforms caused by exposure to 1 M sorbitol occurs in a MAPK-independent manner. Possibly, the hypertonic shock caused by 1 M sorbitol activates an as yet to be identified phosphatase. In any event, both acute and chronic stimulation of the HOG pathway leads to down-regulation of TORC2–Ypk1 signaling, albeit by different mechanisms.

### MAPK phosphorylation of Avo2 alters its subcellular localization

Studies of various integral and peripheral PM proteins in yeast have shown that different components often reside in discrete microdomains (Spira et al. 2012). Many permeases, such as the arginine transporter Can1, colocalize in a membrane compartment called the "MCC" that is congruent with eisosomes (Strádalová et al. 2009), protein-coated PM invaginations whose major constituents include Pil1, an F-BAR (FCH/bin-amphiphysin-rvs) domain-containing protein (Karotki et al. 2011). Initially, it was reported that all of the components of TORC2, including Avo2, reside in a PM microdomain dubbed the "MCT" that is distinct from eisosomes (Berchtold and Walther 2009). However, subsequent work demonstrated that Avo2 partitions between TORC2 and eisosomes (Bartlett et al. 2015). For these reasons, and having established that MAPK phosphorylation does not affect Avo2 stability (Figs. 2A, 5C), to explain how MAPK phosphorylation of Avo2 contributes to down-modulation of TORC2 activity, we examined whether MAPK phosphorylation of Avo2 affected its subcellular localization.

Toward that end, we expressed either Avo2, Avo2[9A], or Avo2[9E] (each tagged at its C terminus with mNeon Green

[mNG]) (Shaner et al. 2013) from its endogenous locus on chromosome XIII in cells coexpressing the eisosome marker Pil1 (tagged at its C terminus with red fluorescent protein [RFP]) (Fradkov et al. 2000) from its endogenous locus on chromosome VII. We document (1) that Avo2-mNG is fully functional (Supplemental Fig. S4) and (2) that both Avo2-mNG and Pil1-RFP displayed a punctate distribution at the PM and that these puncta were often, but not always, congruent (Fig. 5A, left), as we resolved and quantified using CellProfiler and appropriate masking (Supplemental Fig. S5). The same was observed for Avo2[9A]-mNG and Pil1-RFP (Fig. 5A, middle). However, there were significant differences found in the cells expressing Avo2[9E]-mNG and Pil1-RFP. Generally, the Avo2[9E]-containing puncta were smaller and less prominent (Fig. 5A, right), and their overall mean pixel intensity was significantly lower than in cells expressing Avo2 or Avo2[9A] (Fig. 5B) despite the fact that the total amount of Avo2, Avo2[9A], and Avo2[9E] in each of these cells was very similar (Fig. 5C), and the intensity of the Pil1-RFP foci remained essentially unchanged (Fig. 5A,B). Moreover, the decrease in the intensity of the Avo2[9E]-containing PM puncta was the same regardless of whether the Avo2[9E] colocalized with Pil1 (Fig. 5D) or was located at distinct PM sites, presumably with TORC2 (Fig. 5E). Taken together, these findings suggest that MAPK phosphorylation promotes dissociation of Avo2 from either of its two PM sites: eisosomes and TORC2.

To confirm that conclusion for TORC2 specifically by an independent biochemical approach, we immuno-isolated TORC2 (in which the Avo3 subunit carried a 3xFlag epitope tag) from the cells expressing Avo2, Avo2[9A], or Avo2[9E] and analyzed the amount of Avo2 present in these complexes. In the immunoprecipitates obtained using anti-Flag antibodies, equivalent amounts of both Avo3-3xFlag and the Tor2 catalytic subunit were found regardless of whether the cells expressed Avo2, Avo2[9A], or Avo2[9E] (Fig. 5F). Likewise, in cells expressing Tor2-mNG and Avo2, Avo2[9A], or Avo2[9E], there was little or no effect on the amount of Tor2 present in PM-associated TORC2 (Supplemental Fig. S6). However, quantification of the immunoprecipitates showed that, compared with the amount of Avo2 or Avo2[9A] bound to TORC2, the amount of Avo2[9E] was always reduced by 30%–40%. Thus, in agreement with the observations made by fluorescence imaging, these results suggest that MAPK phosphorylation of Avo2 promotes its dissociation from TORC2.

### Discussion

It has been unclear whether TORC2 adjusts its activity by serving itself as a direct sensor of perturbations to the cell envelope or responding to inputs from other stress-sensing pathways (or both). We demonstrated here, for the first time, that Slt2/Mpk1 (mammalian ortholog ERK5/MAPK7) (Truman et al. 2006), the MAPK of the CWI pathway, phosphorylates two subunits of TORC2 (Avo2 and Avo3) and, as a result, down-regulates TORC2

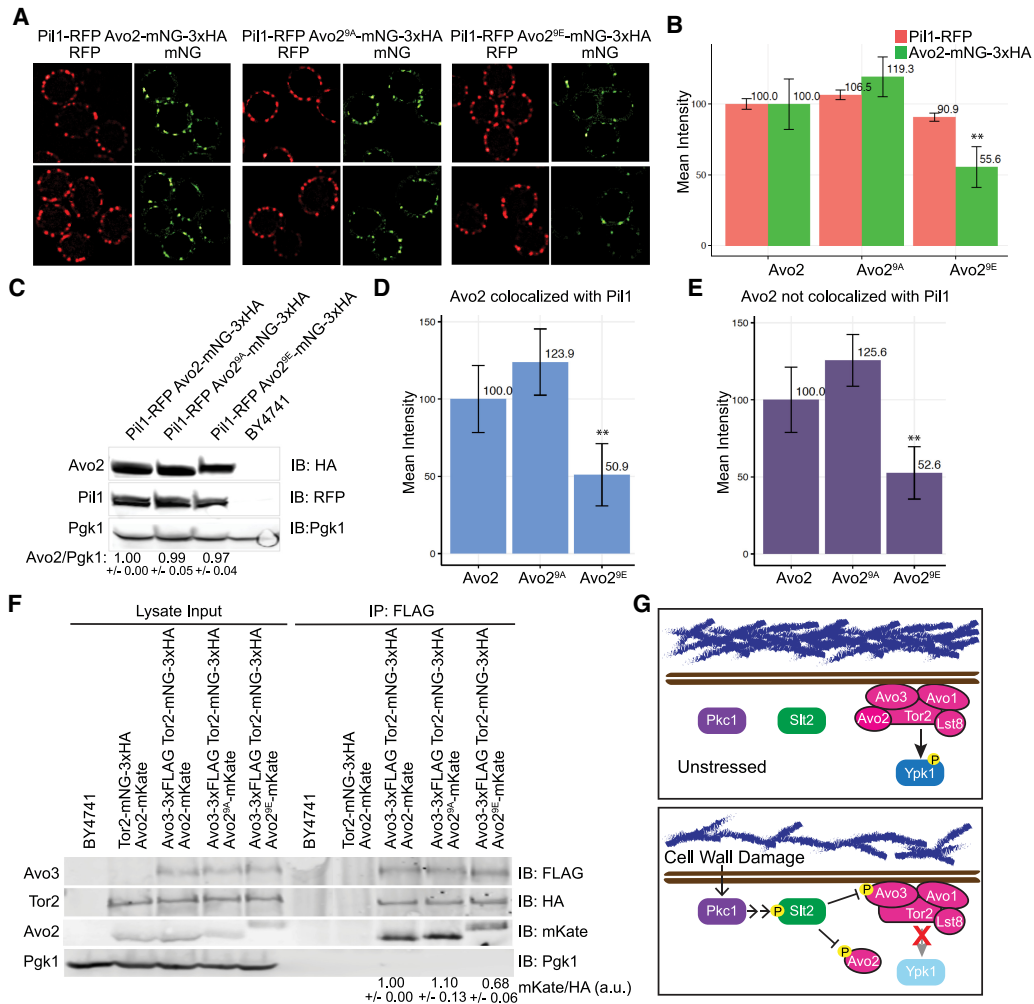

**Figure 5.** Phosphorylation at its MAPK sites displaces Avo2 from the PM. (*A*) Strains yAEA348 (Avo2$^{WT}$-mNG-3xHA Pil1-RFP), yAEA349 (Avo2$^{9A}$-mNG-3xHA Pil1-RFP), and yAEA350 (Avo2$^{9E}$-mNG-3xHA Pil1-RFP) were grown to mid-exponential phase and examined by fluorescence microscopy as described in the Materials and Methods and processed using CellProfiler (as in Supplemental Fig. S3). (*B*) Mean values of the relative pixel intensities of the Pil1-RFP and Avo2-mNG foci in images, as in *A*, were measured using CellProfiler, and those obtained for strain yAEA348 (Avo2$^{WT}$-mNG-3xHA Pil1-RFP) were set to 100%, to which all of the other values were normalized. Error bars represent 95% confidence intervals. (**) $P < 0.0001$. (*C*) The same cells as in *A* were resolved by standard SDS-PAGE and analyzed by immunoblotting. A representative experiment is shown. Values *below* the lanes represent the relative Avo2 to Pgk1 ratio (average of three independent experiments with SEM). (*D*) The mean pixel intensity of the Avo2-mNG foci that colocalized with Pil1-RFP was measured using CellProfiler as in *B*. (**) $P < 0.0001$. (*E*) The mean pixel intensity of the Avo2-mNG foci that did not colocalize with Pil1-RFP was measured using CellProfiler as in *B*. (**) $P < 0.0001$. (*F*) Strains BY4741, YFR589 (Avo2-mKate Tor2-mNG-3xHA), YFR624 (Avo2-mKate Avo3-3xFlag Tor2-mNG-3xHA), YFR626 (Avo2$^{9A}$-mKate Avo3-3xFlag Tor2-mNG-3xHA), and YFR628 (Avo2$^{9E}$-mKate Avo3-3xFlag Tor2-mNG-3xHA) were propagated in YPD, harvested, and lysed, and equal amounts of the resulting extracts were immunoprecipitated with anti-Flag antibody. Proteins in the input extracts and in the immunoprecipitates were resolved by SDS-PAGE and analyzed by immunoblotting. A representative experiment is shown. Values *below* the *right* lanes represent the relative Avo2 to Tor2 ratio (average of three independent experiments with SEM), with the value obtained for strain YFR624 (Avo2-mKate Avo3-3xFlag Tor2-mNG-3xHA) set to 1.00. (*G*) Model showing that under conditions that damage cell wall structure, activation of the CWI pathway exerts feedback that will negatively regulate the growth-promoting functions of TORC2.

phosphorylation of its primary downstream effector, AGC family kinase Ypk1 (mammalian ortholog SGK1) (Casamayor et al. 1999). We also found that modification of Avo2 was sufficient to explain Slt2-mediated inhibition of TORC2, but it is possible that Slt2 phosphorylation of Avo3 also contributes to TORC2 down-modulation. Likewise, all of the other subunits of TORC2 contain potential

MAPK phosphorylation sites; however, unlike for Avo2 and Avo3, phosphorylation at very few of these other candidate sites has been detected in global phosphoproteomic studies.

In any event, the control circuit that we discovered makes physiological sense because under conditions that interfere with cell wall synthesis or damaged cell wall structure, activation of the CWI pathway will exert feedback that will negatively regulate the growth-promoting functions of TORC2 (Fig. 5G). For yeast cell growth, enlargement of the cell wall needs to be tightly coupled to both an increase in cell mass and expansion of the PM. TORC2–Ypk1 signaling controls the processes that maintain adequate PM levels of all of the lipid classes (sphingolipids, glycerolipids, and sterols) needed to sustain growth and viability (Roelants et al. 2017a, 2018). Thus, down-regulation of TORC2 activity upon CWI pathway activation provides a mechanism by which the status of the cell wall can be sensed by TORC2 and thereby the rates of the reactions necessary for PM homeostasis can be adjusted accordingly.

The need for regulatory circuitry in which the state of the cell wall can feed in and influence TORC2 function perhaps explains why a recognizable Avo2 homolog is present throughout the vast majority of the fungal clade—from *S. cerevisiae* and its sensu stricto relatives to divergent yeast species (e.g., *Debaryomyces hansenii* and *Yarrowia lipolytica*) to filamentous fungi (e.g., *Aspergillus nidulans* and *Neurospora crassa*) (see https://portals.broadinstitute.org/cgi-bin/regev/orthogroups/show_orthogroup.cgi?orf=YMR068W)—but is not conserved in multicellular organisms that lack a cell wall. In this same regard, genetic evidence indicates that the Slt2 ortholog (Pmk1) in fission yeast (*Schizosaccharomyces pombe*) has a role in negatively regulating TORC2-mediated activation of the fission yeast Ypk1 ortholog (Gad8) (Cohen et al. 2014), providing further support for conservation of the overall mechanism that we describe here.

Our results show that Avo2 is not only an important site of Slt2-mediated phosphorylation but also necessary for optimal TORC2 activity, especially under stressful conditions. However, the function of Avo2 in TORC2 is not currently known. Within its N-terminal half, Avo2 contains at least five ankyrin repeats (Gaubitz et al. 2016), which are known to mediate protein–protein interactions but, in some cases, also interactions with other classes of biomolecules. Initial dissection of the subunit contacts in yeast TORC2 by coimmunoprecipitation and pull-down assays suggested that Avo2 associates only with Avo1 and Avo3 (Wullschleger et al. 2005). However, recent modeling of a predicted Avo2 structure into the cryo-electron microscopy (cryo-EM) density of yeast TORC2, if interpreted correctly, indicates intimate contact of Avo2 with the HEAT repeat and FAT domain regions of Tor2, much less extensive interaction with Avo3, and no contact with Avo1 (Karuppasamy et al. 2017). Nonetheless, this placement is surface-exposed and quite distant from the active site of the kinase domain in Tor2 and hence does not readily suggest an obvious function for Avo2 (Karuppasamy et al. 2017).

As first detected in global two-hybrid screens, Avo2 also interacts with two other paralogous proteins, Slm1 and Slm2, and the ability of Avo2 to associate with Slm1 and Slm2 also has been observed by both coimmunoprecipitation (Audhya et al. 2004) and in vitro binding assays (Fadri et al. 2005). Slm1 and Slm2 are components of the furrow-like PM invaginations called eisosomes that are coated with the PtdIns4,5$P_2$-binding F-BAR domain-containing proteins Pil1 and Lsp1 (Karotki et al. 2011). Similarly, the central ~190 residues of Slm1 (686 residues) and Slm2 (656 residues) contain a predicted I-BAR (inverse BAR) domain (which is required for their targeting to eisosomes) (Olivera-Couto et al. 2011) and a demonstrated ~110-residue PtdIns4,5$P_2$-binding PH domain near their C-terminal ends (Audhya et al. 2004; Gallego et al. 2010). Absence of Slm1 and Slm2 reportedly disrupts eisosome structure (Kamble et al. 2011).

It has been observed before that each component subunit of TORC2 localizes around the cell periphery in multiple PM-associated "dots" and that these puncta are quite dynamic (Berchtold and Walther 2009), whereas the eisosomes are considered rather static structures (Brach et al. 2011). Moreover, it had been suggested that these two compartments are discrete (Berchtold and Walther 2009). However, as we found and document here, a substantial fraction of Avo2 is readily found in association with the diagnostic eisosome marker Pil1, in agreement with the observations of at least one other group (Bartlett et al. 2015). Furthermore, there is convincing evidence that the Avo2-interacting proteins and eisosome components Slm1 and Slm2 are required for TORC2-mediated phosphorylation of Ypk1 (Berchtold et al. 2012; Niles et al. 2012). However, the model proposed was that, under appropriate stress conditions (including hypotonic shock and sphingolipid depletion), Slm1/2 dissociate from eisosomes, bind Ypk1, and deliver this substrate to the TORC2 compartment (Berchtold et al. 2012; Niles et al. 2012). On the other hand, equally compelling data demonstrate that it is the CRIM element present in Avo1 and its orthologs in fission yeast (Sin1) and mammalian cells (mSIN1) that binds and presents substrates to the catalytic center in TORC2 (Liao and Chen 2012; Tatebe and Shiozaki 2017).

Hence, the findings that we describe here suggest an obverse model; namely, that the role of Avo2 (assuming, as its gene name indicates, that it "associates voraciously" with Tor2 and the other components of TORC2) is to act as an adaptor that, through its association with Slm1/Slm2, is responsible for the dynamic docking of TORC2 at eisosomes. What, then, is the function of the Avo2–Slm1/2 interaction in promoting TORC2 function? As demonstrated by multiple groups (Bultynck et al. 2006; Tabuchi et al. 2006), Slm1/2 also bind tightly Ca$^{+2}$/calmodulin-activated phosphoprotein phosphatase 2B (also known as calcineurin) and do so via canonical so-called PxIxIT motifs located very near their C termini ([673]PNIYIQ[678] in Slm1 and [640]PEFYIE[645] in Slm2). Hence, it is possible that the calcineurin bound to Slm1/2 acts to remove the inhibitory phosphorylations present on Avo2 and other TORC2 subunits, thereby alleviating this negative regulation. Consistent with this model, we found that Avo2[9E], an allele that mimics the permanently phosphorylated state and hence would be immune to calcineurin action, was inhibitory to optimal TORC2 function in vivo, exhibited enhanced dissociation from PM sites, and reduced association with TORC2. Thus, based on the behavior of this phosphomimetic allele, a primary effect of Slt2-mediated phosphorylation is to impede its integration into TORC2 and interfere with Avo2-mediated association of TORC2 with the cell cortex. Whether Slt2 phosphorylation of Avo3 (or other TORC2 subunits) further promotes this dissociation and other ramifications of this model provide fertile ground for further study.

Finally, what we reveal here is one molecular mechanism by which a stress signal restrains a growth-promoting signal. As we document here for yeast, it seems likely that MAPK pathways activated by stress will impede TORC2 function in mammalian cells in a similar manner because continued growth and response to stress are invariably incompatible cellular states.

## Materials and methods

### Construction of yeast strains and growth conditions

Unless indicated otherwise in Table 1, *S. cerevisiae* strains used in this work were all constructed using conventional yeast genetic methods (Sherman 2002). Unless stated otherwise, yeast cultures were grown in standard rich (YP) medium or defined minimal (SC) medium (Sherman 2002) containing 2% glucose/ dextrose and were supplemented with the appropriate nutrients to permit growth of auxotrophs and/or select for plasmids. Cultures were propagated at 30°C, unless indicated otherwise. Induction of genes under the control of *GAL* promoters was carried out either by adding 2% galactose to strains pregrown in the appropriate SC medium with 2% raffinose and 0.2% sucrose followed by incubation over the course of 3 h or, for strains containing the Gal4 human estrogen receptor herpes simplex virus transactivator VP16 fusion protein (Gal4-ER-VP16 [GEV]) (Quintero et al. 2007), by addition of 20 μM β-estradiol (final concentration) followed by incubation over the course of 3 h. For strains containing a protein tagged with an AID, the cells were cultured in YPD or the appropriate SC medium buffered with 50 mM potassium phosphate (pH 6.2), and degradation was induced by addition of a synthetic auxin (1-NAA) (Sigma-Aldrich) for the indicated time periods.

### Plasmids and recombinant DNA methods

Plasmids used in this work (Table 2) were constructed using standard procedures in *Escherichia coli* strain DH5α (Green and Sambrook 2012). All PCR reactions were performed with Phusion high-fidelity DNA polymerase (New England Biolabs, Inc.). Site-directed mutagenesis was performed by using the appropriate mismatch oligonucleotide primers with the QuikChange method (Agilent Technologies, Inc.) according to the manufacturer's instructions. The fidelity of all constructs was verified by nucleotide sequence analysis.

### Cell extract preparation and immunoblotting

Samples of exponentially-growing cells were harvested by brief centrifugation and stored at −80°C. Cell pellets were thawed on ice and lysed in 150 μL of 1.85 M NaOH and 7.4% β-mercaptoethanol. Proteins were precipitated by the addition of 150 μL of 50% trichloroacetic acid for 10 min on ice. Precipitated proteins were pelleted by centrifugation and washed twice with ice-cold acetone. Protein pellets were solubilized in 5% SDS in 0.1 M Tris base, adjusted to a final concentration of 0.025 $A_{600\,nm}$ per microliter, mixed with one-fifth volume of 5× SDS sample buffer, boiled for 10 min, and then resolved by SDS-PAGE. Phospho-proteins were resolved by Phos-tag SDS-PAGE (Kinoshita et al. 2009): Phosphorylated Ypk1-myc was resolved in 8% acrylamide gels containing 35 μM Phos-tag affinity reagent (Wako Chemicals USA, Inc.) and 70 μM $MnCl_2$, phosphorylated Avo2-3xFlag was resolved by in 8% acrylamide gels containing 17.5 μM Phos-tag reagent and 35 μM $MnCl_2$, and phosphorylated Avo3-3xFlag was resolved in 8% acrylamide gels containing 20 μM Phos-tag reagent and 40 μM $MnCl_2$.

For immunoblotting, proteins in gels were transferred electrophoretically to a nitrocellulose membrane, incubated with Odyssey blocking buffer (Li-Cor Biosciences, Inc.) diluted 1:1 with PBS or TBS, and then probed by addition of the appropriate primary antibody at the indicated dilution: mouse anti-myc mAb 9E10 (1:100; Monoclonal Antibody Facility, Cancer Research Laboratory, University of California at Berkeley), mouse anti-Flag M2 mAb (1:10,000; Sigma-Aldrich), mouse anti-HA.11 mAb (1:1000; BioLegend), rabbit polyclonal anti-RFP (1:5000; Rockland Immunochemicals), rabbit polyclonal anti-tRFP (1:1000 to detect mKate; Evrogen), or rabbit polyclonal anti-Pgk1 (1:30,000; this laboratory, prepared as described in Baum et al. 1978). Activated (dually phosphorylated) Hog1 was detected using rabbit anti-p38 MAPK phospho-Thr180/phospho-Tyr182 mAb (1:1000; Cell Signaling Technology), and activated (dually phosphorylated) Slt2 was detected using rabbit anti-p44/42 (Erk1/2) MAPK phospho-Thr202/phospho-Tyr204 mAb (1:1000; Cell Signaling Technology). After rinsing the membranes, filter-bound immune complexes were detected with an appropriate infrared dye-labeled secondary antibody—CF770-conjugated goat anti-mouse IgG (Biotium), IRDye800CW-conjugated goat anti-rabbit IgG (Li-Cor), or IRDye680RD-conjugated goat anti-mouse IgG (Li-Cor) —that was diluted 1:10,000 in 1:1 Odyssey blocking buffer::PBS (or TBS) containing 0.1% Tween-20 and 0.02% SDS and visualized using an infrared imaging system (Odyssey CLx, Li-Cor).

### Slt2 in vitro kinase assay

To purify activated Slt2, 1 L each of yeast strain CGA84 cotransformed with $P_{GAL1}$-$PKC1^*$ and either pKL63 ($P_{GAL1}$-Slt2$^{WT}$-TAP) or pKL64 ($P_{GAL1}$-Slt2$^{KD}$-TAP) was grown to mid-exponential phase at 30°C. Expression was induced by addition of 20 μM β-estradiol followed by incubation for 3 h. The cells were harvested, washed once in ice-cold TAP-B buffer containing phosphatase and protease inhibitors (200 mM NaCl, 1.5 mM MgOAc, 1 mM DTT, 2 mM $NaVO_4$, 10 mM NaF, 10 mM Na-PPi, 10 mM β-glycerol phosphate, 1× complete protease inhibitor [Roche], 50 mM Tris-HCl at pH 7.5), resuspended in 4 mL of TAP-B buffer, and flash-frozen as droplets in liquid $N_2$. The cells were ruptured cryogenically in a Mixer Mill MM301 (Retsch). The resulting lysate was thawed on ice, diluted with 8 mL of TAP-B buffer, and clarified by centrifugation at 15,000g for 20 min. The resulting clarified extract was subjected to centrifugation at 78,000g for 1 h, and the resulting supernatant fraction was adjusted to a final concentration of 0.15% NP-40 using an appropriate volume of a 10% NP-40 stock. Slt2-TAP in the detergent-containing soluble fraction was captured by binding to IgG-agarose resin (GE Healthcare). After washing the resin extensively with TAP wash buffer (200 mM NaCl, 1.5 mM MgOAc, 1 mM DTT, 0.01% NP-40, 2 mM $NaVO_4$, 10 mM NaF, 10 mM Na-PPi, 10 mM β-glycerol phosphate, 50 mM Tris-HCl at pH 7.5), it was washed with Protease 3C buffer (200 mM NaCl, 1.5 mM MgOAc, 1 mM DTT, 0.01% NP-40, 10% glycerol, 2 mM $NaVO_4$, 10 mM NaF, 10 mM Na-PPi, 10 mM β-glycerol phosphate, 50 mM Tris-HCl at pH 7.5). To elute the bound Slt2-TAP, the washed resin was resuspended in 500 μL of Protease 3C buffer containing 60 U of PreScission protease (GE Healthcare). After incubation for 5 h at 4°C, the resin was

**Table 1.** *S. cerevisiae strains used in this study*

| Strain | Genotype | Source or reference |
|---|---|---|
| BY4741 | *MATa his3Δ1 leu2Δ0 met15Δ0 ura3Δ0* | Research Genetics, Inc. |
| BY4742 | *MATα his3Δ1 leu2Δ0 lys2Δ0 ura3Δ0* | Research Genetics, Inc. |
| JTY5336 | BY4741 *trp1::URA3* | This study |
| YFR549 | BY4741 *slt2Δ::KanMX trp1::URA3* | This study |
| YFR538-A | BY4742 *hog1Δ::KanMX trp1::URA3* | This study |
| YFR560 | BY4741 *kss1Δ::CgHIS3 trp1::URA3* | This study |
| YFR559 | BY4741 *slt2Δ::KanMX hog1Δ::KanMX trp1::URA3* | This study |
| YFR567 | BY4742 *slt2Δ::KanMX kss1Δ::CgHIS3 trp1::URA3* | This study |
| YFR564 | BY4741 *hog1Δ::KanMX kss1Δ::CgHIS3 trp1::URA3* | This study |
| YFR617 | BY4741 Avo3-3xFlag::KanMX *trp1::URA3* | This study |
| JTY5473 | BY4741 *sln1ts-4::KanMX* | Costanzo et al. 2010 |
| yKL15 | BY4741 Sln1-6xHA::HygR TIR1::*HIS3* | This study |
| yKL18 | BY4741 Sln1-AID*-6xHA::HygR TIR1::*HIS3* | This study |
| yKL20 | BY4741 *lys2Δ0* Sln1-AID*-6xHA::HygR *hog1Δ::KanMX* TIR1::*HIS3* | This study |
| yKL16 | BY4741 Sln1-AID*-6xHA *slt2Δ::URA3* TIR1::*HIS3* | This study |
| yKL22 | BY4741 Sln1-AID*-6xHA *hog1Δ::KanMX slt2Δ::URA3* TIR1::*HIS3 MET15* | This study |
| JTY7318 | BY4742 *avo2Δ::KanMX* | Research Genetics, Inc. |
| yKL32 | BY4742 Avo2-3xFlag::*URA3* | This study |
| YFR181 | BY4742 Pil1-RFP *LYS2 met15Δ0* | This study |
| YFR528 | BY4742 Avo2(T144A T219A S233A S240A S249A T310A S315A T330A S333A)::*URA3* | This study |
| yKL31 | BY4742 Avo2(T144E T219E S233E S240E S249E T310E S315E T330E S333E)::*URA3* | This study |
| yKL33 | BY4742 Avo2(T144A T219A S233A S240A S249A T310A S315A T330A S333A)-3xFlag::*URA3* | This study |
| yKL34 | BY4742 Avo2(T144E T219E S233E S240E S249E T310E S315E T330E S333E)-3xFlag::*URA3* | This study |
| yEPS2 | BY4741 Pil1(S163D S230D T233D::*URA3* | This study |
| YFR641 | BY4741 Avo2(T144E T219E S233E S240E S249E T310E S315E T330E S333E)::*URA3* Pil1(S163D S230D T233D::*URA3* | This study |
| yAEA348 | BY4741 Pil1-RFP Avo2-mNG-3xHA::*URA3* | This study |
| yAEA349 | BY4741 Pil1-RFP Avo2(T144A T219A S233A S240A S249A T310A S315A T330A S333A)-mNG-3xHA::*URA3* | This study |
| yAEA350 | BY4741 Pil1-RFP Avo2(T144E T219E S233E S240E S249E T310E S315E T330E S333E)-mNG-3xHA::*URA3* | This study |
| yAEA301-A[a] | BY4742 Tor2-mNG-3xHA | |
| YFR589 | BY4741 Avo2-mKate::SpHIS5 Tor2-mNG-3xHA | This study |
| YFR624 | BY4741 Avo2-mKate::SpHIS5 Tor2-mNG-3xHA Avo3-3xFlag::*KanMX* | This study |
| YFR626 | BY4741 Avo2(T144A T219A S233A S240A S249A T310A S315A T330A S333A)-mKate::SpHIS5 Tor2-mNG-3xHA Avo3-3xFlag::*KanMX* | This study |
| YFR628 | BY4741 Avo2(T144E T219E S233E S240E S249E T310E S315E T330E S333E)-mKate::SpHIS5 Tor2-mNG-3xHA Avo3-3xFlag::*KanMX* | This study |
| CGA84 | *MATa leu2Δ1::GEV::NATMX pep4Δ::HIS3 prb1Δ1.6R ura3-52 trp1-1 lys2-801a leu2Δ1 his3Δ200 can1 GAL* | Alvaro et al. 2014 |

[a]For strain yAEA301-A, the coding sequence for an mNG-3xHA cassette was inserted in-frame between Asn321 and Thr322 of the *TOR2* ORF on chromosome XI, constructed using a CRISPR–Cas9-based method. In brief, BY4742 was transformed with *URA3*-marked pRS316-*GAL_prom*-Cas9 (pGF-V789) (Finnigan and Thorner 2016), expression of Cas9 was induced in a resulting transformant by addition of galactose (2% final concentration), and, after 6 h, the cells were cotransformed with *HIS3*-marked pRS423-sgRNA-Tor2 (pAEA272) and a *TOR2*-mNG-3xHA-*TOR2* repair cassette generated and amplified by overlap extension PCR using appropriate DNA templates and synthetic oligonucleotide primers. The desired (viable) transformants were selected on SCD – Ura – His plates and then colony-purified by streaking on SD + Ura + His plates containing 5-FOA (to select for derivatives that lost pGF-V789 and pAEA272). The single-guide RNA (sgRNA) encoded by pRS423-sgRNA-Tor2 (pAEA272) was −463ATTTTGGCGCCAGTGA TAAGTCG−441 (the PAM sequence is underlined) directed against the corresponding sequence within the *TOR2* promoter (where +1 is the first base of the initiator ATG of the *TOR2* ORF). In the upstream primer used to generate the *TOR2*-mNG-3xHA-*TOR2* repair cassette DNA, to prevent recognition by the sgRNA, the sequence in and around the PAM was mutated to −446TAGTCG−441. The proper in-frame insertion of the mNG-3xHA-coding sequence in the *TOR2* ORF was verified by colony PCR and DNA sequencing.

removed by centrifugation, and the Protease 3C in the resulting supernatant faction was removed by incubation with a slurry of glutathione-Sepharose beads (GE Healthcare) followed by centri-fugation. The affinity-purified Slt2 in the resulting final supernatant solution was diluted into kinase assay buffer (125 mM potassium acetate, 5% glycerol, 12 mM MgCl2, 0.5 mM EGTA,

**Table 2.** *Plasmids used in this study*

| Plasmid | Description | Source or reference |
|---|---|---|
| pRS315 | *CEN, LEU2*, vector | Sikorski and Hieter 1989 |
| pFR246 | pRS315 Ypk1(T51A T71A T504A S644A T662A)-myc | Leskoske et al. 2017 |
| pKL1 | pRS315 Avo2-3xFlag | This study |
| pKL2 | pRS315 Avo2(T144A T219A S233A S240A S249A T310A S315A S330A S333A)-3xFlag | This study |
| pFR234 | pRS315 Ypk1(D242A)-myc | Roelants et al. 2017b |
| yEplac112 | *2 µm, TRP1*, vector | Gietz and Sugino 1988 |
| Pkc1* (pJT5660) | YEplac112 P$_{GAL1}$-Pkc1(R398A R405A R406A) | Martín et al. 2000 |
| pJEN12 (Pkc1*KD) | YEplac112 P$_{GAL1}$-Pkc1(R398A R405A R406A D949A) | This study |
| pAEA376 | YEplac112 P$_{GAL1}$-Pkc1(R398A R405A R406A)-myc | This study |
| pJEN13 | YEplac112 P$_{GAL1}$-Pkc1(R398A R405A R406A D949A)-myc | This study |
| BG1805 | *2 µm, URA3*, P$_{GAL1}$, C-terminal tandem affinity (TAP) tag vector | Open Biosystems, Inc. |
| pKL63 | BG1805 Slt2 | This study |
| pKL64 | BG1805 Slt2(K54R) | This study |
| pGEX6P-1 | GST tag, bacterial expression vector | GE Healthcare, Inc. |
| pKL16 | pGEX6P-1 Avo2 | This study |
| pKL17 | pGEX6P-1 Avo2(T144A T219A S233A S240A S249A T310A S315A S330A S333A) | This study |
| pKL81 | pGEX6P-1 Avo3(1–100) | This study |
| pKL82 | pGEX6P-1 Avo3(1–100)(S11A T24A T28A S50A S84A S87A) | This study |

2 mM DTT, 1 mM PMSF, 0.1 mM chymostatin, 4 mM AEBSF, 12.5 mM β-glycerol phosphate, 1 mM NaVO$_4$, 40 mM Tris-HCl at pH 7.5).

Substrates to be tested were purified as recombinant proteins by fusion to the C terminus of GST in appropriate pGEX vectors and expressed in *E. coli* BL21(DE3)pLysS. One liter of cultures was grown at 30°C to mid-exponential phase and then induced with 0.5 mM IPTG for 3 h at 30°C. The bacterial cells were harvested by centrifugation and lysed, and the GST fusion protein of interest was adsorbed to glutathione-agarose beads (Prometheus Laboratories Inc.), which were washed and resuspended in kinase assay buffer. The same volume of either affinity-purified Slt2$^{WT}$ or Slt2$^{KD}$ was added to the resuspended beads or an equivalent volume of kinase assay buffer, and reaction was initiated by the addition of 2 µCi of [γ-$^{32}$P]ATP and unlabeled ATP to 100 µM final concentration. After incubation for 30 min at 30°C, reactions were terminated by addition of 5× SDS-PAGE sample buffer followed by boiling for 10 min. The resulting reaction products were resolved by SDS-PAGE and analyzed by Coomassie blue staining and autoradiography using a phosphorimager (Typhoon, GE Healthcare).

### Immunoprecipitation of TORC2

One-hundred milliliters of cultures of yeast strains expressing Avo3-3xFlag Tor2-mNG-3xHA and either Avo2-mKate (YFR624), Avo2$^{9A}$-mKate (YFR626), or Avo2$^{9E}$-mKate (YFR628) was grown in YPD to mid-exponential phase, harvested by centrifugation, and frozen in liquid nitrogen. Cells pellets were resuspended in 500 µL of 2× TNEG buffer (100 mM Tris at pH 7.6, 300 mM NaCl, 20% glycerol, 0.24% Tergitol, 2 mM EDTA, 0.1 mM PMSF, 1× Roche complete protease inhibitor tablet [Roche]). Glass beads were added to the meniscus of the cell suspension, and lysis was achieved by vigorous vortex mixing for 3 min. After lysis, 2× vol of 1× TNEG buffer was added, and then the lysate was clarified by centrifugation at 1000*g* for 5 min at 4°C followed by a second centrifugation at 2000*g* for 10 min. TORC2 complexes marked by the tightly bound Avo3-3xFlag subunit were collected by immunoabsorption to 40 µL of mouse anti-Flag affinity gel (Biotool) equilibrated in 1× TNEG buffer. The resin was washed four times with 1 mL of 1× TNEG buffer and resuspended in 2× urea buffer.

### Fluorescence microscopy and image analysis

Fluorescence microscopy was performed using an Elyra PS.1 structured illumination microscope (Carl Zeiss AG) equipped with a 100× plan-apo 1.46NA TIRF objective, a main focus drive of the AxioObserver Z1 stand, and a WSB PiezoDrive 08 controlled by Zen, and images were recorded using a 512×512 (100-nm × 100-nm pixel size) EM-CCD camera (Andor Technology). To visualize Avo2 or Tor2 tagged with mNG (excitation λ$_{max}$ 506 nm; emission λ$_{max}$ 517), cell samples were excited with an argon laser at 488 nm at 2.3% power (100 mW), and emission was captured in a 495- to 550-nm window using a bandpass filter; for Pil1 tagged with RFP/DsRed (excitation λ$_{max}$ 558 nm; emission λ$_{max}$ 583), excitation was at 561 nm at 2.3% power (100 mW), and emission was monitored in a 570- to 620-nm window using a different bandpass filter. Images (average of eight scans; 300 msec/scan) were analyzed using Fiji (Schindelin et al. 2012). To avoid changes in image quality due to occasional fluctuations in laser intensity, all panels shown in any given figure represent experiments performed on the same day and scaled and adjusted identically for brightness using Fiji (Schindelin et al. 2012). For quantitative automated analysis of fluorescence intensity at the PM, CellProfiler was used (Carpenter et al. 2006). To train CellProfiler to apply the appropriate mask and quantify the signal, a corresponding pipeline was created, which was adapted from prior software (Bray et al. 2015; Chong et al. 2015). Prior to loading into the CellProfiler pipeline, cell images were segmented manually using Fiji (Schindelin et al. 2012). To avoid any selection bias, every cell visible in the bright-field image in a frame from any sample (except those out of focus) was chosen. All plots and statistical analyses in this study were performed with the R statistical analysis package (http://www.R-project.org).

## Acknowledgments

We thank Victor J. Cid and Maria Molina (University of Madrid Compluteuse) for the Pkc1* plasmid, Steve Ruzin and Denise Schichnes (Biological Imaging Facility, University of California at Berkeley) for their invaluable advice about fluorescence microscopy, and all members of the Thorner laboratory for helpful discussions. We apologize in advance to any investigator whose work was not cited due to the limitation on the total number of allowed references imposed by the editorial policy of this journal. This work was supported by National Institutes of Health (NIH) Predoctoral Traineeship GM07232 and a University of California at Berkeley MacArthur and Lakhan-Pal Graduate Fellowship to K. L.L., Erwin Schroedinger Fellowship J3787-B21 from the Austrian Science Fund to A.E.-A., Marie Sklodowska-Curie Individual Fellowship GA 750835 from the European Commission to C.M.A., and NIH R01 research grant GM21841 to J.T. This work was also aided in part by NIH S10 Equipment Grant OD018136 (to Steven E. Ruzin, Director, University of California at Berkeley Biological Imaging Facility) for a Zeiss Elyra S1 structured illumination microscope.

*Author contributions*: K.L.L. and F.M.R. designed, executed, and analyzed experiments and drafted the manuscript. A.E.-A. designed, executed, and analyzed experiments and conducted fluorescence microscopy. C.M.A. set up the specific CellProfiler pipeline to quantify the fluorescent intensities and performed all of the statistical analyses. E.P.S. and J.M.H. provided technical assistance with several experiments. J.T. designed and analyzed experiments and revised the manuscript.

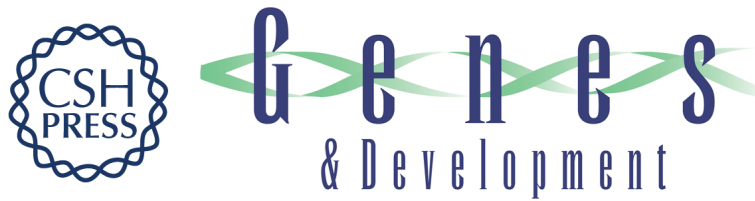

# Phosphorylation by the stress-activated MAPK Slt2 down-regulates the yeast TOR complex 2

Kristin L. Leskoske, Françoise M. Roelants, Anita Emmerstorfer-Augustin, et al.

| | |
|---|---|
| **Supplemental Material** | http://genesdev.cshlp.org/content/suppl/2018/11/23/gad.318709.118.DC1 |
| | Published online November 26, 2018 in advance of the full issue. |
| **Creative Commons License** | This article, published in *Genes & Development*, is available under a Creative Commons License (Attribution 4.0 International), as described at http://creativecommons.org/licenses/by/4.0/. |
| **Email Alerting Service** | Receive free email alerts when new articles cite this article - sign up in the box at the top right corner of the article or **click here.** |

# Supplemental Figure 1

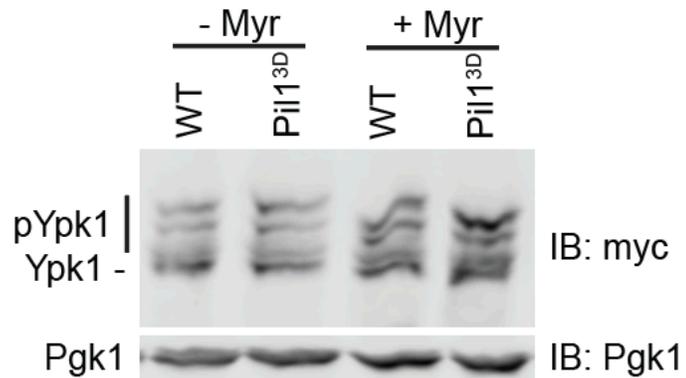

Fig. S1. **TORC2 function is not affected by phospho-mimetic Pil1 allele.** Wild-type (BY4742) or Pil1$^{3D}$ (yEPS2) strains were cultured to mid-exponential phase in selective minimal medium and then treated with vehicle (methanol) or 1.25 µM myriocin for 2 h. After harvesting, whole-cell lysates were prepared, resolved by SDS-PAGE, and analyzed as described in Materials and Methods.

# Supplemental Figure 2

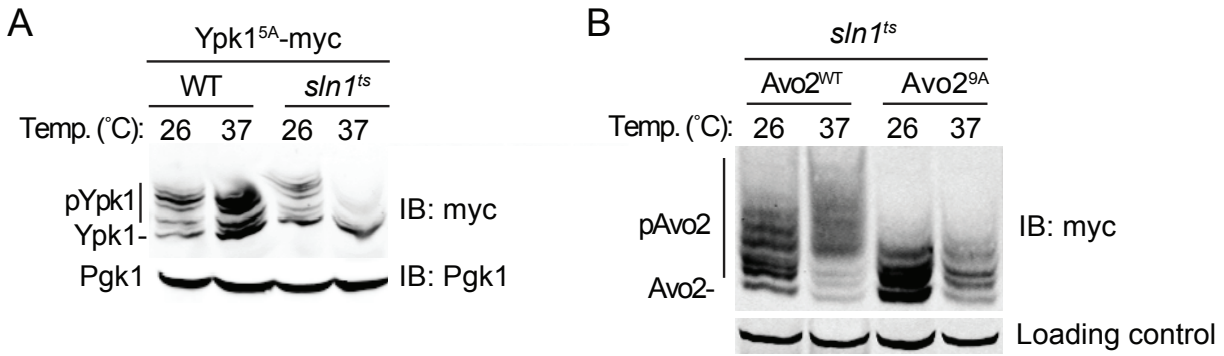

**Fig. S2. TORC2 function is down-regulated after Sln1 inactivation.** (A) Wild-type cells (BY4741) and an otherwise isogenic *sln1ts* derivative (JTY5473) expressing Ypk1$^{5A}$-myc from its native promoter on a *CEN* plasmid (pFR246) were propagated at 26˚C to mid-exponential phase and then a portion of the culture was maintained at 26˚C and another portion shifted to 37˚C. After 2 h at the respective temperatures, samples of the separate cultures were harvested, lysed and Ypk1 phosphorylation analyzed by Phos-tag SDS-PAGE and immunoblotting. Pgk1 was the loading control. (B) Strain JTY5473 (*sln1ts*) expressing either Avo2-3xFLAG (pKL1) or Avo2$^{9A}$-3xFLAG (pKL2) were grown to mid-exponential phase and then either kept at 26˚C or shifted to 37˚C. After 2 h, Avo2 phosphorylation was analyzed as in (A).

# Supplemental Figure 3

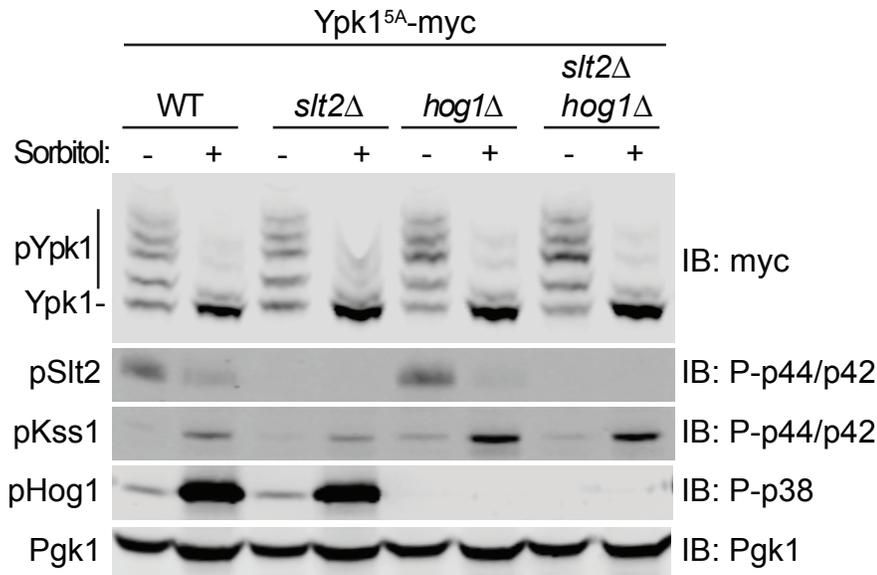

Fig. S3. **Challenge with a high exogenous sorbitol concentration rapidly diminished TORC2-mediated Ypk1 phosphorylation**. Strains JTY5336 (*HOG1*[+] *SLT2*[+]), YFR549 (*slt2Δ*), YFR538A (*hog1Δ*), and YFR559 (*slt2Δ hog1Δ*), each expressing Ypk1[5A]-myc from its native promoter on a *CEN* plasmid (pFR246), were grown to mid-exponential phase, then diluted into fresh SCD-Leu medium in the absence (-) or presence (+) of sorbitol (1 M final concentration). After 5 min, the cells were harvested, lysed and the resulting extracts subjected to Phos-tag SDS-PAGE to resolve Ypk1 phosphorylation and to standard SDS-PAGE to resolve the other proteins, and analyzed by immunoblotting, as described in Materials and Methods.

**Supplemental Figure 4**

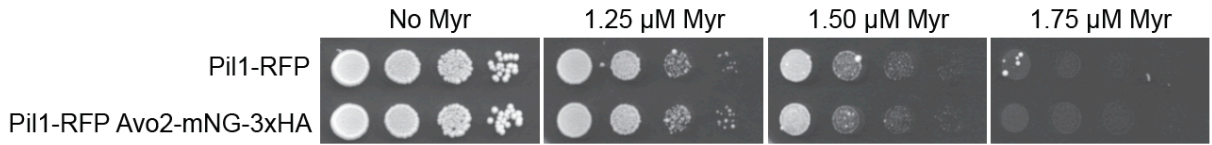

Fig. S4. **Avo2-mNG-3xHA is fully biologically functional.** Overnight cultures of otherwise isogenic Pil1-RFP (yFR181) and Pil1-RFP Avo2-mNG-3xHA (yAEA348) strains were adjusted to $A_{600nm}$ = 0.1, spotted undiluted and in a series of 10-fold serial dilutions on YPD plates either lacking or containing myriocin at the indicated concentrations, incubated at 30°C for 3 d, and then imaged.

# Supplemental Figure 5

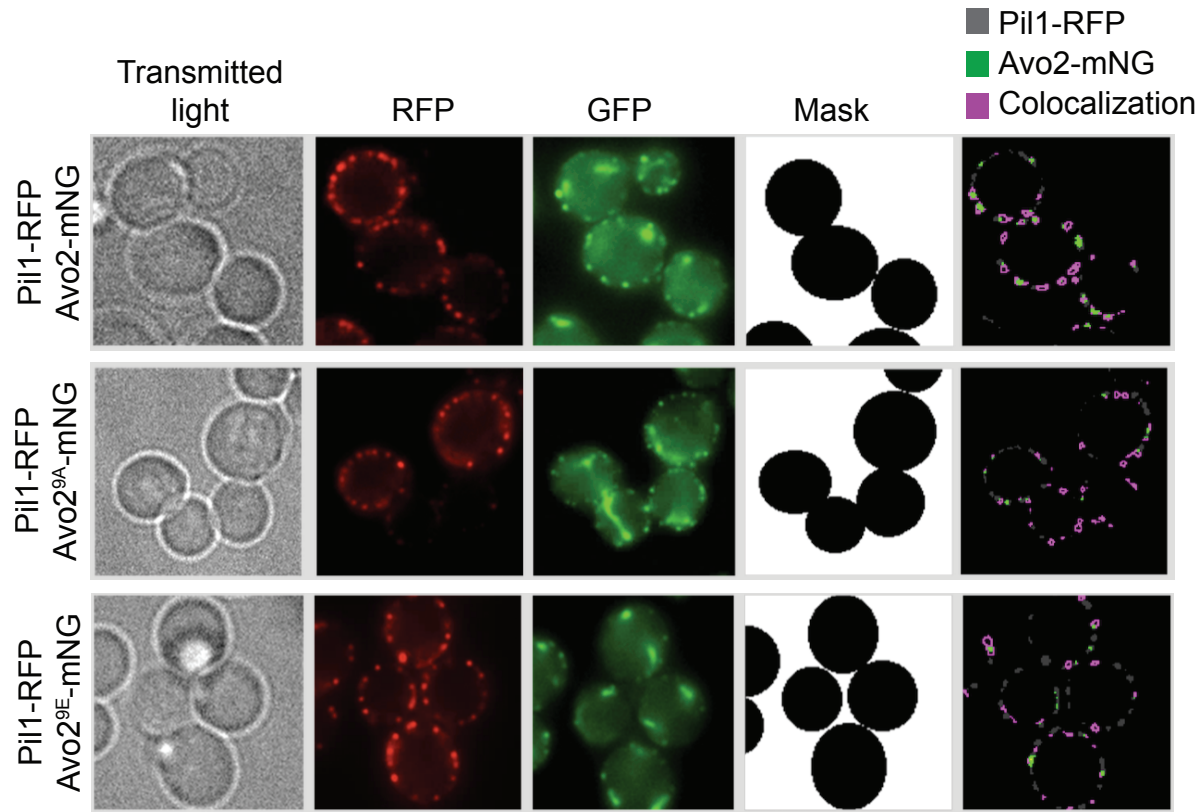

Fig. S5. **Illustration of the masking used to quantify fluorescent images of the cell periphery with CellProfiler**. The images and other data shown in Fig. 5 were obtained using a custom CellProfiler pipeline, as described in Materials and Methods. The blob-like or thread-like cytosolic structures observed in the GFP channel are due to intrinsic autofluorescence of the mitochondria.

# Supplemental Figure 6

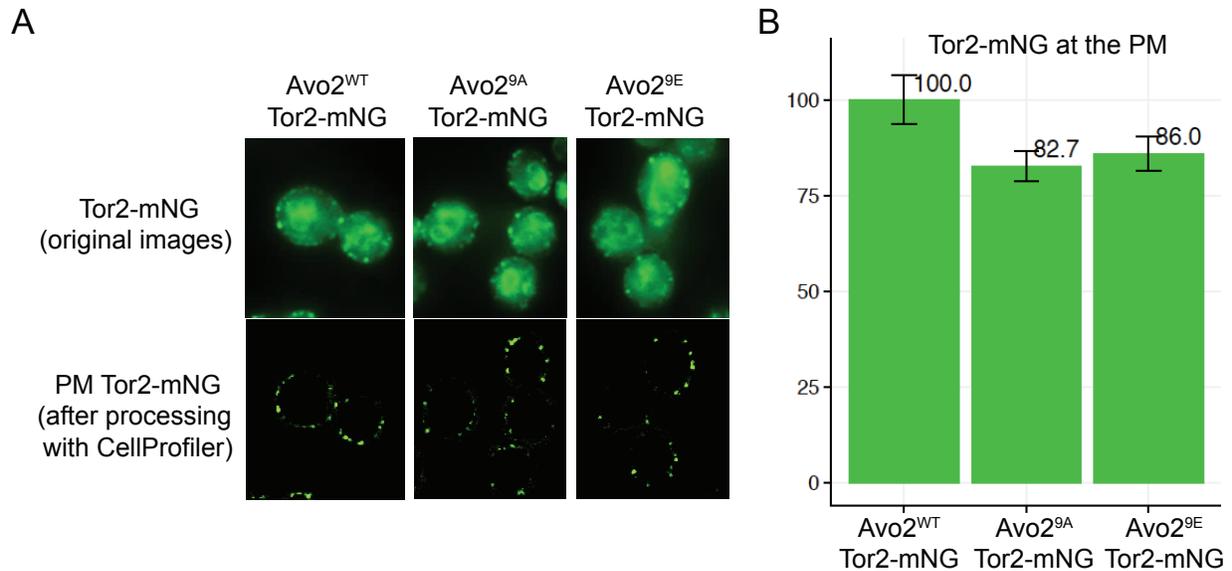

Fig. S6. **The state of Avo2 phosphorylation does not affect Tor2 localization.** (A) Strains YFR624 (Avo2-mKate Tor2-mNG-3xHA Avo3-3xFLAG), YFR626 (Avo2$^{9A}$-mKate Tor2-mNG-3xHA Avo3-3xFLAG) and YFR628 (Avo2$^{9E}$-mKate Tor2-mNG-3xHA Avo3-3xFLAG) were grown to mid-exponential phase and examined by fluorescence microscopy as described in Materials and Methods and processed using CellProfiler (as in Fig. S3). (B) Mean values of the relative pixel intensities of the Tor2-mNG foci in images, as in (A), were measured using CellProfiler, and those obtained for strain YFR624 (Avo2-mKate Tor2-mNG-3xHA Avo3-3xFLAG) were set to 100%, to which all the other values were normalized. Error bars represent 95% confidence intervals.